# Importance of the atom-pair bond in metallic alloying

T. Rajasekharan[1,*] and V. Seshubai[2,**]

[1]*Defence Metallurgical Research Laboratory, Kanchanbagh P.O. Hyderabad 500 058, India,*

[2]*School of Physics, University of Hyderabad, Hyderabad 500 019, India.*

Rajasekharan and Girgis reported that binary systems with intermetallic compounds of a particular crystal structure form a straight line on a map using Miedema's parameters. In this paper, the universality of that observation is examined. Observations from a study of 143 binary systems that crystallize in six different crystal structures at $AB_3$ composition are discussed. Prediction of concomitant and mutually exclusive structure types in binary metallic phase diagrams, and of phase transitions among different structure types, has been demonstrated. This behaviour is unexpected because Miedema's parameters are isotropic in nature and structural energies are generally assumed to be small. We argue in this paper that each point on the map stands for the energy of an unlike atom-pair (A-B) bond, with the bond energy remaining nearly the same at all compositions in the phase diagram. This argument is confirmed by comparing the nearest-neighbour (A-B) bond lengths for the compounds of the structure types $CaCu_5$ and CsCl, when concomitant with $MgCu_2$ structure type. This fact leads to an important conclusion that one can define a bond energy for the metallic unlike-atom-pair bond as is usually done for a conventional chemical bond.

## I. INTRODUCTION

A large amount of experimental data exists in the literature on metallic alloy phase diagrams. A study of such data using the average properties or lumped constants for atoms such as electronegativity and valence can be expected to help frame laws which will be useful in predicting the outcome of experiments/ processes in laboratories and industries, without resorting to extensive calculations every time.

Miedema's semi--empirical theory[1-4] for the heats of formation ($\Delta$H) of metallic alloys is considered highly successful. It allows the prediction of the signs of $\Delta$H with almost 100% accuracy. Miedema's equation is essentially empirical in nature with scope for reinterpretation of the physical meanings of the parameters $\phi$ and $N$. $\phi$ is called the 'work function' and $N$, the 'electron density at the boundary of the Wigner-Seitz cell' of an element. The prediction of the crystal structures adopted by intermetallic compounds in binary systems has also interested scientists for a long time and they have worked essentially using structural maps with different pseudo-potential-derived or empirically derived parameters as coordinates[5-9]. Intermetallic compounds with different crystal structures were separated into different regions on such maps. Some of the above approaches have been reviewed by Machlin[10] and others[11,12].

Rajasekharan and Girgis studied[13,14] the behaviour of intermetallic compounds on structural maps with $\Delta\phi$ and $\Delta N^{1/3}$ as coordinates (hereafter referred to as $RG$ maps), where $\Delta$ denotes the difference in the quantities between two alloying elements. Earlier attempts[15] to construct structural maps with the absolute values $|\Delta\phi|$ and $|\Delta N^{1/3}|$ as coordinates were unsuccessful. It can be observed from binary phase diagrams that intermetallic compounds with a particular crystal structure are often accompanied by compounds with certain other crystal structures at other compositions. It is also observed that compounds with certain crystal structure types do not occur together in the same phase diagram. For instance in the 250 binary systems in which Laves phases occur, $SiCr_3$ type compounds do not occur and in the 87 binary systems in which $SiCr_3$ type compounds occur, Laves phases do not occur. The only exceptions are the Mo-Be system and the Ir-Zr system. $BeMo_3$ and $IrZr_3$ are $SiCr_3$ type. $ZrIr_2$ is $MgCu_2$ type and $MoBe_2$ is $MgZn_2$ type. Savitskii and Gribulya[16] have made similar observations which they proposed as a guideline for looking for new representatives of the σ -phase. In the binary systems in which the σ -phase exists, they pointed out that $SiCr_3$, Laves ( $MgZn_2$ type) and CsCl type compounds usually coexist. Rajasekharan and Girgis had shown that their maps could predict concomitant and mutually exclusive structure types in binary phase diagrams. This was a surprising result because Miedema's theory is supposed to be isotropic. Also, the structural energies are generally assumed to be small. Hence the observed systematics in the behaviour of binary systems on the $RG$ map is not anticipated from the current understanding of the alloying of metals. The present paper evolved during our efforts to understand the above observations.

In the first part of this paper, we discuss various features exhibited by some of the structure types on the ($\Delta\phi$, $\Delta N^{1/3}$) map. We then argue that the observations made on the map prove that the points on the map represent the energy of an unlike atom-pair bond A−B in an alloy, which remains nearly the same at all compositions in a binary system. We, therefore, propose that the metallic A−B bond is like a conventional chemical bond where the energy of the bond remains more or less the same irrespective of the functional groups attached to the atoms[17]. This conclusion is new and is rather unanticipated, and we confirm it further by comparing the first nearest neighbour unlike atom distances in structure types that coexist in binary systems.

## II. MIEDEMA'S THEORY AND RAJASEKHARAN – GIRGIS MAPS

Following the early work of Axon[18] and Mott[19], Miedema et al.[1-4] developed a semi-empirical model to calculate $\Delta H$ of binary metallic alloys. Miedema's equation for $\Delta H$, for a compound at equiatomic composition, is given below:

$$\Delta H = \left[-\Delta\phi^2 + \frac{Q}{P}\left(\Delta N^{1/3}\right)^2 - \frac{R}{P}\right] \quad \ldots(1).$$

$\phi$ and $N$ are the work function and the 'electron density at the boundary of the Wigner-Seitz cells' of the elements and their numerical values were obtained from Pauling's electronegativity and experimental bulk modulus of the elements respectively. Their values were then adjusted within uncertainties in their determination to predict the signs of $\Delta H$ of more than 500 binary systems, using Eq. (1), with nearly 100% accuracy. Considerable amount of empiricism was involved in developing Miedema's equation. It was reported that the choice of $\left(\Delta N^{1/3}\right)^2$ in Eq. (1), rather than $(\Delta N)^2$ which followed from Miedema's theoretical arguments, gave a better fit to the experimental data[2]. $P$ and $Q$ then become universal constants for all elemental combinations with $Q/P$ = 9.4 Volts/(d.u.)$^{1/3}$. The value of $Q/P$ was obtained from a fit of the experimental data. The term $R/P$, increasing with the valence of the $p$-metals, was introduced to correctly reproduce the empirically observed data on the signs of $\Delta H$ of transition metal - $p$-metal combinations. $R/P$ is zero for other combinations. Miedema et al. suggested that $R/P$ has its origin in '$d - p$ hybridisation'[1]. Eq. (1) was multiplied by a function $f(c)$ to arrive at the numerical values of $\Delta H$ for compounds of different stoichiometries[2]. $f(c)$ was a function of the elemental concentrations and volumes and its form was chosen to reproduce the experimentally observed variations[2]. Though the *signs* of $\Delta H$ were predicted very accurately by Eq. (1), the magnitudes of $\Delta H$ could be estimated only to a much lower accuracy[20,21].

We have recently studied[22] all the structure types in which at least 5 intermetallic compounds crystallise (96 structure types in all with ~3000 intermetallic compounds), and found that the observations made by Rajasekharan and Girgis[13] are true for all of them. The data on the structures of intermetallic compounds that we have studied was collected from standard sources[23-25]. Various structure types dealt with in this paper are discussed briefly in Table I and a full description can be found in Pearson's book[26]. Our aim was to see whether for all the known structure types, the linear dependence between $\Delta\phi$ and $\Delta N^{1/3}$ is true. We also studied the characteristics of the lines, their distribution on the ($\Delta\phi$, $\Delta N^{1/3}$) map, the elemental combinations that make up the points in different regions of the map and examined as to whether it is universally true that the $RG$ maps with ($\Delta\phi$, $\Delta N^{1/3}$) coordinates can predict concomitant and mutually exclusive structure types in phase diagrams.

Corresponding to a binary system A-B, there can be two symmetrically opposite points on the $(\Delta\phi,\ \Delta N^{1/3})$ map: $\left(\left(N_A^{1/3} - N_B^{1/3}\right), (\phi_A - \phi_B)\right)$ and $\left(\left(N_B^{1/3} - N_A^{1/3}\right), (\phi_B - \phi_A)\right)$. For a compound at the composition $A_m B_n$ $(n > m)$, we plot $(\phi_A - \phi_B)$ versus $\left(N_A^{1/3} - N_B^{1/3}\right)$. For the 1:1 composition AB, we plot $(\phi_A - \phi_B)$ versus $\left(N_A^{1/3} - N_B^{1/3}\right)$, with the element A identified as the one to the left of B in the periodic table. We call, for brevity, the straight line formed by the binary systems in which compounds of a particular structure type occur, on the $(\Delta\phi,\ \Delta N^{1/3})$ map as the '$RG$ line' and the line that is obtained by plotting $(\phi_B - \phi_A)$ versus $\left(N_B^{1/3} - N_A^{1/3}\right)$ as the '$Inverse\ RG$ line' of the structure type. We note that the composition at which a structure type occurs does not enter into deciding its coordinates on the $RG$ map; thus each point on the map represents a binary system and not a particular intermetallic compound.

We observed that out of the 96 structure types studied, 88 formed single straight lines on the $RG$ map, the rest formed more than one line[22]. In this paper, we show in Fig. 1, the $RG$ lines of a number of structure types occurring at the composition $AB_3$ in binary metallic systems. The lines of slope $\pm\sqrt{Q/P}$ through the origin correspond to $\Delta H = 0$ in Eq. (1) and the hyperbolae corresponding to $R/P = 2.3$ mark the boundary between the regions of negative and positive $\Delta H$ for $t-p$ compounds. All the elements for which Miedema's parameters are available[1] are considered on the maps. The rule that the binary systems with a particular type of structure fall on a straight line is seen to be followed for all the structure types studied, with practically no compounds as exceptions. All the $RG$ lines have a positive slope and the observed slopes are nearly $+\sqrt{(Q/P)}$ as for the $\Delta H = 0$ line. Some of the observed characteristics of the lines are given in Table I.

In a given binary system, we can predict from Fig. 1, using the elemental properties $\phi$ and $N$, the structures that will be adopted by intermetallic compounds at compositions 1:3 and 3:1, from among those considered. There are situations where two or more lines overlap in some regions: for instance the $AsNa_3$ and $BiF_3$ lines in Fig. 1. In the region of overlap, there are many $AsNa_3$ type compounds which transform to the $BiF_3$ type and vice versa as a function of temperature: The compounds $BiK_3$ and $BiRb_3$ which are $AsNa_3$ type at lower temperatures transform to $BiF_3$ type at higher temperatures[25]. $SbLi_3$ has $BiF_3$ type structure at lower temperatures and transforms to $AsNa_3$ type at higher temperatures[25]. Pearson has pointed out the similarities of these two structure types[27] that the minority element is surrounded by the majority element in the form of two tetrahedra with strong $sp^3$ bonding. Such structure types are closely related in their energy and in such cases it is difficult to predict which of the two structures would be adopted by a given $AB_3$ compound. Closely related structure types have their $RG$ lines nearly coincident; e.g. the Laves phases $MgCu_2$, $MgZn_2$ and $MgNi_2$ types[13]. Many representatives of these structure types often transform from one to the other as a function of temperature. We also observe that when the $RG$ lines do not overlap, there is no instance of structural transformation between corresponding structure types as a function of temperature. For instance, an examination of compiled data[25] shows that not a single $SiCr_3$ type compound would transform into $CoGa_3$ type as a function of temperature and vice versa.

$RG$ lines of structure types occurring at various stoichiometry can be plotted together on the same map. The overlap and lack of overlap between the lines corresponding to different structure types on the $RG$ map can be used to predict concomitants and mutually exclusive structure types in phase diagrams with great accuracy. In Fig. 1, we have included the $RG$ line corresponding to the $MgCu_2$ type compounds occurring at the composition $AB_2$. The figure tells, for instance, that if an $MgCu_2$ type compound occurs at the composition $AB_2$ in a binary system A−B, an $AsNa_3$ type (or $CoGa_3$ type or $SiCr_3$ type) compound will not occur at composition $AB_3$. The question whether an $AsNa_3$ type phase would occur at composition $A_3B$ is decided by whether the inverse $RG$ line of $MgCu_2$ type coincides with the $AsNa_3$ line.

The $\Delta\phi$ versus $\Delta N^{1/3}$ plot corresponding to a structure type is observed to be linear irrespective of the composition at which it occurs, with slope around $+\sqrt{Q/P}$. Its existence region is bounded by the $\pm\sqrt{Q/P}$ lines through the origin (and by the hyperbolae at $R/P$ = 2.3 in the case of the $t-p$ compounds). All compounds, irrespective of their compositions, occur in the region predicted by Eq. (1) to have a negative $\Delta H$.

An examination of metallic phase diagrams[25] revealed that there are 22 structure types which coexist with the $MgCu_2$ type compounds in various binary systems. We have studied the behaviour of those structure types on the $RG$ map. In Fig. 2, we show the $RG$ lines of $Si_3Mn_5$ (hP16, 145 representatives) and $BaAl_4$ (tI10, 18 representatives) structure types along with the $RG$ and the inverse $RG$ lines of the $MgCu_2$ type (cF24, 181 representatives). The lines shown are the best fit lines obtained using the individual data sets, and the big cluster of points (seen black online) are the binary systems in which $MgCu_2$ type compounds co-exist with the $Si_3Mn_5$ type or $BaAl_4$ type compounds. The concomitant compounds are found to occur near the regions of intersection of the lines.

## III. SIGNIFICANCE OF SHORTEST UNLIKE ATOM-PAIR BOND

Miedema's equation predicts the signs of the heats of formation of *both liquid and solid alloys*[3] with ~100% accuracy suggesting that long range order has practically no role in deciding the signs of the heats of formation of metallic alloys. We observe that binary systems having a particular structure type occur on a straight line on the $RG$ map with a slope $\approx +\sqrt{Q/P}$, and that this is true for all the structure types. When we consider all the compounds belonging to *different* structure types, the only common feature among them will be the interaction on a line joining the A and B atoms, i.e. the pair-wise interaction between the A and B atoms. Hence our observations on the $RG$ map suggest that $\Delta\phi \propto \Delta N^{1/3}$ with the proportionality constant $\approx +\sqrt{Q/P}$, for the A−B pair-wise interaction.

In addition, we note that to accurately predict concomitant and mutually exclusive structure types in phase diagrams from the $RG$ maps, no input needs to be made regarding the compositions at which the structure types occur. In a binary system A-B, irrespective of whether we are considering a compound occurring at AB, $AB_2$, $AB_3$ or in general any $A_mB_n$, the structures of the concomitant phases are predicted by a knowledge of which $RG$ lines or inverse $RG$ lines pass through the point corresponding to the binary system on the $(\Delta\phi,\ \Delta N^{1/3})$ map. This means that the energy represented by a point on the $RG$ map (and given by Eq. (1)) is the energy of the pair-wise interaction between A and B atoms or is the A−B bond energy, which remains the same for all stoichiometry and structures involved. The conclusion that the A−B bond energy is a constant at different compositions is similar to the idea in conventional chemistry that the bond energies for a particular type of bond, say an S−H bond, is found to be approximately constant in different molecules[17] containing that type of bond. The above conclusion marks an important deviation from the way we look at the metallic bonds. Pauling considered[28] charge transfer in alloy systems and used an assumption along similar lines to arrive at the volume of an A−B ion-pair in the binary system Ca−Pb with the compounds $Ca_2Pb$, $Ca_5Pb_3$, CaPb (tP4), CaPb (cP4) and $CaPb_3$. He assumed that after equal number of Ca and Pb ions form bonds, the extra Ca and Pb atoms have the effective volumes of the elementary substances (i.e. they are not affected by the bonding process). Using the observed values of the mean atomic volume in the alloy, he then calculated the volumes of the $Ca^{+} - Pb^{-}$ ion pair in all the intermetallic compounds in the Ca−Pb system and found them to be nearly the same. Similar observations were reported by him[28] in the case of Co−Zr and Co−Ga systems.

In order to understand the physical picture involved, we look at the crystal structures adopted by intermetallic compounds. They are characterised[29] by high symmetry, large connectivity and high ligancy (≥ 8). The first nearest neighbour A−B bonds are usually ≥ 8 in number and are of equal length. For instance, in the cubic $MgCu_2$ compound (Fig. 3(a)), each Mg atom is surrounded[30] by 12 Cu atoms at a distance of 2.922 Å. The second shortest A−B bond length in $MgCu_2$ is 4.578 Å, about 57% longer. We propose that the heat of formation of these compounds is primarily due to covalent bonds resonating among the large number of equivalent nearest neighbour A−B pairs, with the farther bonds contributing to a much lesser extent. A similar picture with a large number of equivalent nearest neighbour A−B bonds emerges if one examines other structure types like $SiCr_3$ type, $MoSi_2$ type etc. Calculations of effective pair potentials have been reported which show that the nearest neighbour pair interaction energy is more important than further-neighbour interactions[31]. The hypothesis that the energy of the nearest neighbour A−B bond plays an important role in deciding the heat of formation of alloys, and the conclusion drawn from the $RG$ map that the A−B bond energy is nearly the same at different compositions in a binary system, would imply that the nearest neighbour A−B bond lengths would be nearly the same in different intermetallic compounds of the same binary system. We show in Table II that it is indeed so in *all* the 43 binary systems in which $MgCu_2$ type (Fig. 3(a)) and $CaCu_5$ type (Fig 3(b)) compounds coexist. The ratio of the shortest A−B bond lengths is uniformly close to 1.05. Such an effect is also demonstrated for *all* the 33 binary systems in which $MgCu_2$ and CsCl type compounds co-exist (Table III). These observations support the physical picture that we have assumed.

## IV. SUMMARY

By a study of all crystal structure types in which intermetallic compounds crystallise, we have observed that binary systems having intermetallic compounds with a particular crystal structure fall on a straight line on a ($\Delta\phi,\ \Delta N^{1/3}$) map, where $\phi$ and $N$ are Miedema's parameters. The map can be used to predict possible structures adopted by intermetallic compounds in a binary system by observing $RG$ lines corresponding to which structure types pass through the point representing that binary system on the map.

We have identified Eq. (1) of Miedema et al. with the energy of the nearest neighbour A−B bond. Each point on the $(\Delta\phi,\ \Delta N^{1/3})$ map corresponds to a single value of energy for the A−B bond. All the RG lines corresponding to structure types concomitant in a binary system would then pass through the point corresponding to the energy of the A−B bond in that system, thus explaining the ability of the RG maps to predict concomitant and mutually exclusive structure types in metallic binary systems. Crystal structures adopted by intermetallic compounds are characterised by high ligancy and symmetry, and the nearest neighbour unlike atom-pair bond is very significant in deciding the heat of formation of the intermetallic compounds. When the energy of the nearest neighbour atom-pair bond goes positive, the compound would become unstable irrespective of the composition of the compound. This explains as to why Miedema's theory would predict the signs of the heats of formation of an alloy accurately, but not necessarily its magnitude.

## ACKNOWLEDGEMENTS

TR thanks DMRL, Hyderabad, India for permission to publish this paper. VS thanks UGC, India for research funding under UPE and CAS programs and CMSD facility at the University.

*trajasekharan@gmail.com , ** seshubai@gmail.com

**Table I** The structure types discussed in the present paper and the characteristics of their $RG$ lines

| S. No | Proto-type | Pearson symbol | Space group | No. of rep. | Details of linear fit on the $RG$ lines: $\Delta\phi = m\,\Delta N^{1/3} + c$ | | |
|---|---|---|---|---|---|---|---|
| | | | | | $m$ | $c$ | $R$ |
| 1 | CsCl | cP2 | Pm3m | 196 | 3.27 | -0.53 | 0.86 |
| 2 | CrB | oC8 | Cmcm | 53 | 2.7 | -0.78 | 0.95 |
| 3 | $MgCu_2$ | cF24 | Fd3m | 181 | 2.9 | -0.44 | 0.95 |
| 4 | $MoSi_2$ | tI6 | I4/mmm | 33 | 3.8 | 0.997 | 0.96 |
| | | | | | 3.8 | -0.68 | 0.95 |
| 5 | $SiCr_3$ | cP8 | $Pm\bar{3}n$ | 61 | 2.55 | 0.9 | 0.93 |
| 6 | $AsNa_3$ | hP8 | $P6_3/mmc$ | 20 | 2.58 | 0.6 | 0.96 |
| 7 | $BiF_3$ | cF16 | Fm 3 m | 28 | 2.95 | 0.55 | 0.98 |
| 8 | $CoAs_3$ | cI32 | $Im\bar{3}$ | 11 | 2.78 | -0.46 | 0.95 |
| 9 | $CoGa_3$ | tP16 | $P\bar{4}n2$ | 8 | 2.35 | 0.03 | 0.77 |
| 10 | $TiAl_3$ | tI8 | I4/mmm | 15 | 2.96 | -0.94 | 0.93 |
| 11 | $BaAl_4$ | tI10 | I4/mmm | 14 | 2.01 | -0.67 | 0.97 |
| 12 | $CaCu_5$ | hP6 | P6/mmm | 65 | 2.65 | -0.63 | 0.96 |
| 13 | $Si_3Mn_5$ | hP16 | $P6_3/mcm$ | 41 | 2.4 | 0.96 | 0.97 |

**TABLE II.** The shortest bond lengths (d in Å) in the $MgCu_2$ type and $CaCu_5$ type compounds occurring in the same binary system are compared. The ratio of the bond lengths is ≈ 1.05 on an average.

| Binary system | $CaCu_5$ type a (Å) | c(Å) | $d_1$(Å) | $MgCu_2$ tpe a (Å) | $d_2$ (Å) | RATIO $d_2/d_1$ |
|---|---|---|---|---|---|---|
| Ba-Pd | 5.49 | 4.34 | 3.170 | 7.95 | 3.296 | 1.04 |
| Ba-Pt | 5.51 | 4.34 | 3.181 | 7.92 | 3.284 | 1.03 |
| Ca-Ni | 4.93 | 3.93 | 2.846 | 7.26 | 3.010 | 1.06 |
| Ce-Co | 4.93 | 4.02 | 2.846 | 7.16 | 2.968 | 1.04 |
| Ce-Ni | 4.88 | 4.01 | 2.817 | 7.22 | 2.993 | 1.06 |
| Ce-Pt | 5.37 | 4.38 | 3.100 | 7.73 | 3.205 | 1.03 |
| Dy-Co | 4.93 | 3.99 | 2.846 | 7.19 | 2.981 | 1.05 |
| Dy-Ni | 4.87 | 3.97 | 2.812 | 7.16 | 2.968 | 1.06 |
| Dy-Rh | 5.14 | 4.29 | 2.968 | 7.49 | 3.105 | 1.05 |
| Er-Co | 4.89 | 4.00 | 2.820 | 7.16 | 2.966 | 1.05 |
| Er-Ni | 4.86 | 3.97 | 2.804 | 7.13 | 2.955 | 1.05 |
| Er-Rh | 4.86 | 3.97 | 2.804 | 7.44 | 3.086 | 1.10 |
| Gd-Co | 4.97 | 3.97 | 2.869 | 7.25 | 3.006 | 1.05 |
| Gd-Ni | 4.91 | 3.97 | 2.835 | 7.20 | 2.985 | 1.05 |
| Gd-Rh | 5.17 | 4.31 | 2.985 | 7.56 | 3.134 | 1.05 |
| Ho-Co | 4.88 | 4.01 | 2.818 | 7.17 | 2.974 | 1.06 |
| Ho-Ni | 4.87 | 3.97 | 2.812 | 7.14 | 2.959 | 1.05 |
| La-Ir | 5.39 | 4.20 | 3.112 | 7.69 | 3.188 | 1.02 |
| La-Ni | 5.02 | 3.99 | 2.898 | 7.39 | 3.064 | 1.06 |
| La-Pt | 5.39 | 4.38 | 3.112 | 7.78 | 3.225 | 1.04 |
| Lu-Ni | 4.83 | 3.97 | 2.791 | 7.06 | 2.929 | 1.05 |
| Nd-Co | 5.02 | 3.98 | 2.898 | 7.22 | 2.993 | 1.03 |
| Nd-Ir | 5.32 | 4.33 | 3.074 | 7.60 | 3.151 | 1.03 |
| Nd-Ni | 4.93 | 3.96 | 2.846 | 7.27 | 3.014 | 1.06 |
| Nd-Pt | 5.35 | 4.39 | 3.089 | 7.69 | 3.188 | 1.03 |
| Pr-Co | 5.01 | 3.98 | 2.893 | 7.31 | 3.031 | 1.05 |
| Pr-Ni | 4.96 | 3.98 | 2.864 | 7.28 | 3.018 | 1.05 |
| Pr-Pt | 5.35 | 4.39 | 3.089 | 7.65 | 3.172 | 1.03 |
| Pu-Ni | 4.87 | 3.97 | 2.812 | 7.14 | 2.960 | 1.05 |
| Pu-Pt | 5.26 | 4.39 | 3.037 | 7.63 | 3.163 | 1.04 |
| Sc-Ni | 4.74 | 3.76 | 2.737 | 6.93 | 2.873 | 1.05 |
| Sm-Co | 5.00 | 3.96 | 2.888 | 7.27 | 3.012 | 1.04 |
| Sm-Ni | 4.93 | 3.98 | 2.844 | 7.23 | 2.997 | 1.05 |
| Sr-Pd | 5.41 | 4.42 | 3.123 | 7.82 | 3.242 | 1.04 |
| Sr-Pt | 5.39 | 4.37 | 3.112 | 7.77 | 3.221 | 1.04 |
| Th-Co | 4.95 | 3.98 | 2.856 | 7.21 | 2.989 | 1.05 |
| Tb-Ni | 4.89 | 3.97 | 2.826 | 7.16 | 2.968 | 1.05 |
| Tb-Rh | 5.13 | 4.29 | 2.964 | 7.49 | 3.106 | 1.05 |
| Th-Ir | 5.33 | 4.27 | 3.077 | 7.66 | 3.177 | 1.03 |
| Yb-Ni | 4.84 | 3.96 | 2.796 | 7.09 | 2.941 | 1.05 |
| Y-Co | 4.93 | 3.99 | 2.846 | 7.22 | 2.993 | 1.05 |
| Y-Ni | 4.88 | 3.97 | 2.817 | 7.19 | 2.981 | 1.06 |
| Y-Rh | 5.14 | 4.29 | 2.968 | 7.50 | 3.109 | 1.05 |

**TABLE III.** The shortest bond lengths in the $MgCu_2$ type and CsCl type compounds occurring in the same binary system are compared. The ratio of the bond lengths is ≈ 1.06 on an average.

| System | CsCl type | | $MgCu_2$ type | | RATIO |
|---|---|---|---|---|---|
| | a (Å) | $d_3$ (Å) | a (Å) | $d_4$ (Å) | $d_4/d_3$ |
| Be-Ti | 2.94 | 2.546 | 6.45 | 2.674 | 1.05 |
| Ca-Pd | 3.52 | 3.048 | 7.67 | 3.180 | 1.04 |
| Dy-Rh | 3.40 | 2.944 | 7.49 | 3.105 | 1.05 |
| Er-Rh | 3.36 | 2.910 | 7.44 | 3.086 | 1.06 |
| Gd-Rh | 3.44 | 2.979 | 7.56 | 3.134 | 1.05 |
| Hf-Co | 3.16 | 2.737 | 6.91 | 2.865 | 1.05 |
| Ho-Ir | 3.38 | 2.927 | 7.49 | 3.105 | 1.06 |
| Ho-Rh | 3.38 | 2.927 | 7.43 | 3.078 | 1.05 |
| Lu-Ir | 3.32 | 2.875 | 7.45 | 3.090 | 1.07 |
| Lu-Rh | 3.33 | 2.884 | 7.42 | 3.075 | 1.07 |
| Mg-Ce | 3.91 | 3.386 | 8.73 | 3.619 | 1.07 |
| Mg-Cd | 3.82 | 3.308 | 8.55 | 3.545 | 1.07 |
| Mg-La | 3.97 | 3.438 | 8.79 | 3.644 | 1.06 |
| Mg-Nd | 3.86 | 3.343 | 8.66 | 3.590 | 1.07 |
| Mg-Pr | 3.88 | 3.360 | 8.69 | 3.603 | 1.07 |
| Mg-Sm | 3.85 | 3.334 | 8.62 | 3.574 | 1.07 |
| Pu-Ru | 3.36 | 2.910 | 7.48 | 3.101 | 1.07 |
| Sc-Al | 3.45 | 2.988 | 7.58 | 3.143 | 1.05 |
| Sc-Co | 3.15 | 2.728 | 6.92 | 2.869 | 1.05 |
| Sc-Ir | 3.21 | 2.776 | 7.35 | 3.046 | 1.10 |
| Sc-Ni | 3.17 | 2.745 | 6.93 | 2.873 | 1.05 |
| Sm-Rh | 3.47 | 3.005 | 7.54 | 3.126 | 1.04 |
| Tb-Rh | 3.42 | 2.962 | 7.49 | 3.106 | 1.05 |
| Ti-Co | 2.99 | 2.590 | 6.69 | 2.774 | 1.07 |
| Tm-Ir | 3.35 | 2.901 | 7.48 | 3.100 | 1.07 |
| Tm-Rh | 3.36 | 2.910 | 7.42 | 3.075 | 1.06 |
| Yb-Ir | 3.35 | 2.901 | 7.48 | 3.099 | 1.07 |
| Yb-Rh | 3.35 | 2.901 | 7.43 | 3.081 | 1.06 |
| Y-Ir | 3.41 | 2.953 | 7.52 | 3.118 | 1.06 |
| Y-Rh | 3.41 | 2.953 | 7.50 | 3.109 | 1.05 |
| Zr-Co | 3.19 | 2.763 | 6.95 | 2.881 | 1.04 |
| Zr-Ir | 3.32 | 2.875 | 7.35 | 3.047 | 1.06 |
| Zr-Zn | 3.34 | 2.893 | 7.39 | 3.064 | 1.06 |

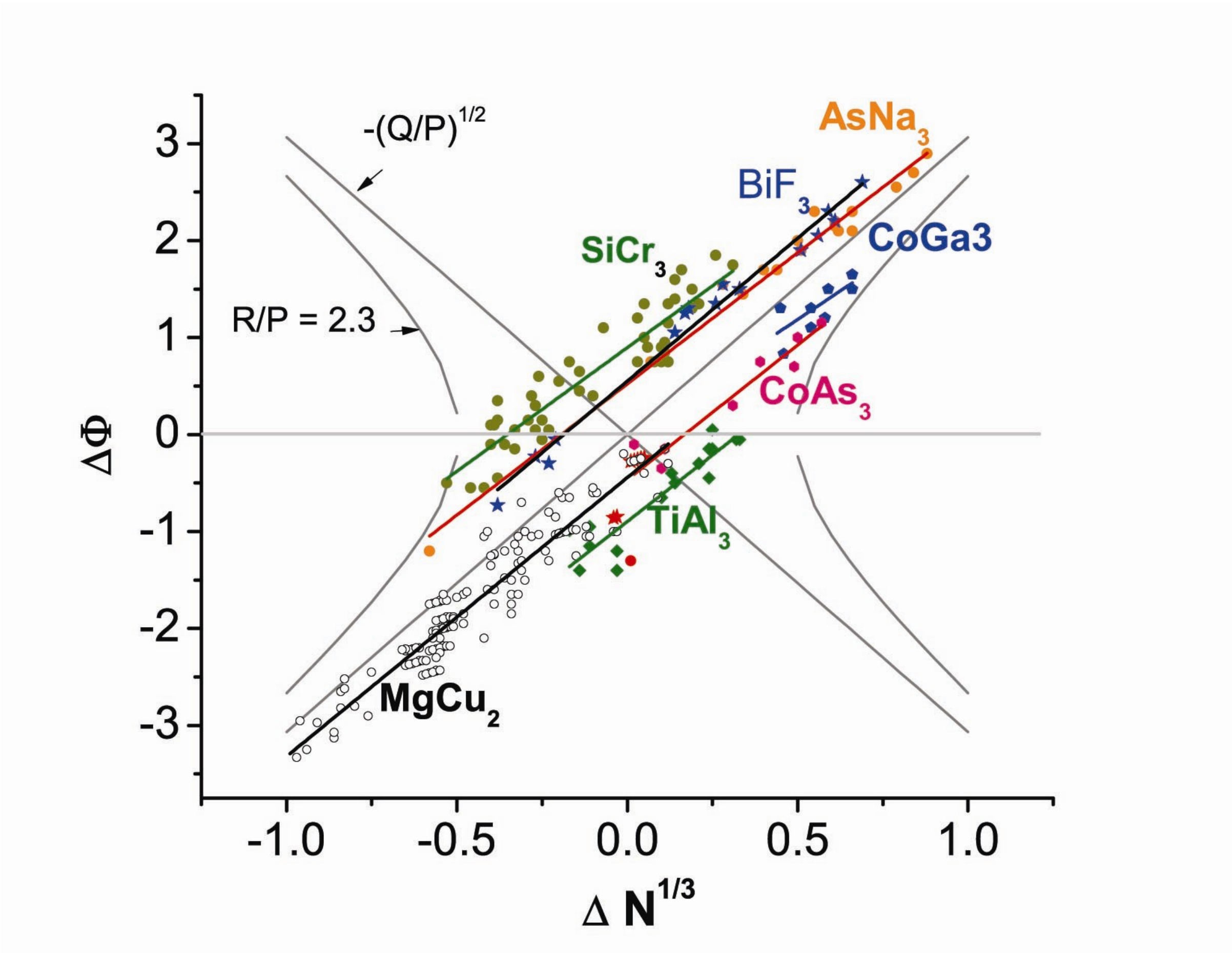

**FIG. 1** (color online). Examples of Rajasekharan-Girgis ($RG$) lines of some structure types occurring at the stoichiometry $AB_3$ are shown along with that of the $MgCu_2$ type occurring at composition $AB_2$. We see that they all form straight lines with positive slopes nearly equal to that of the line that demarcates the regions of positive and negative heats of formation. By observing whether a binary system occurs on the line corresponding to a structure type or not, it is possible to predict whether that structure type can occur in the binary system. Also the structure types coexisting in a particular binary system have their $RG$ lines passing through the point corresponding to that binary system. We can predict, for instance, from the above figure that if an $MgCu_2$ type compound occurs in a binary system at composition $AB_2$, an $SiCr_3$ type compound will not occur at composition $AB_3$. In the overlap region of $AsNa_3$ type and $BiF_3$ type, the compounds $BiK_3$, $BiRb_3$ and $SbLi_3$ show structural transformations between the two structure types as a function of temperature[25].

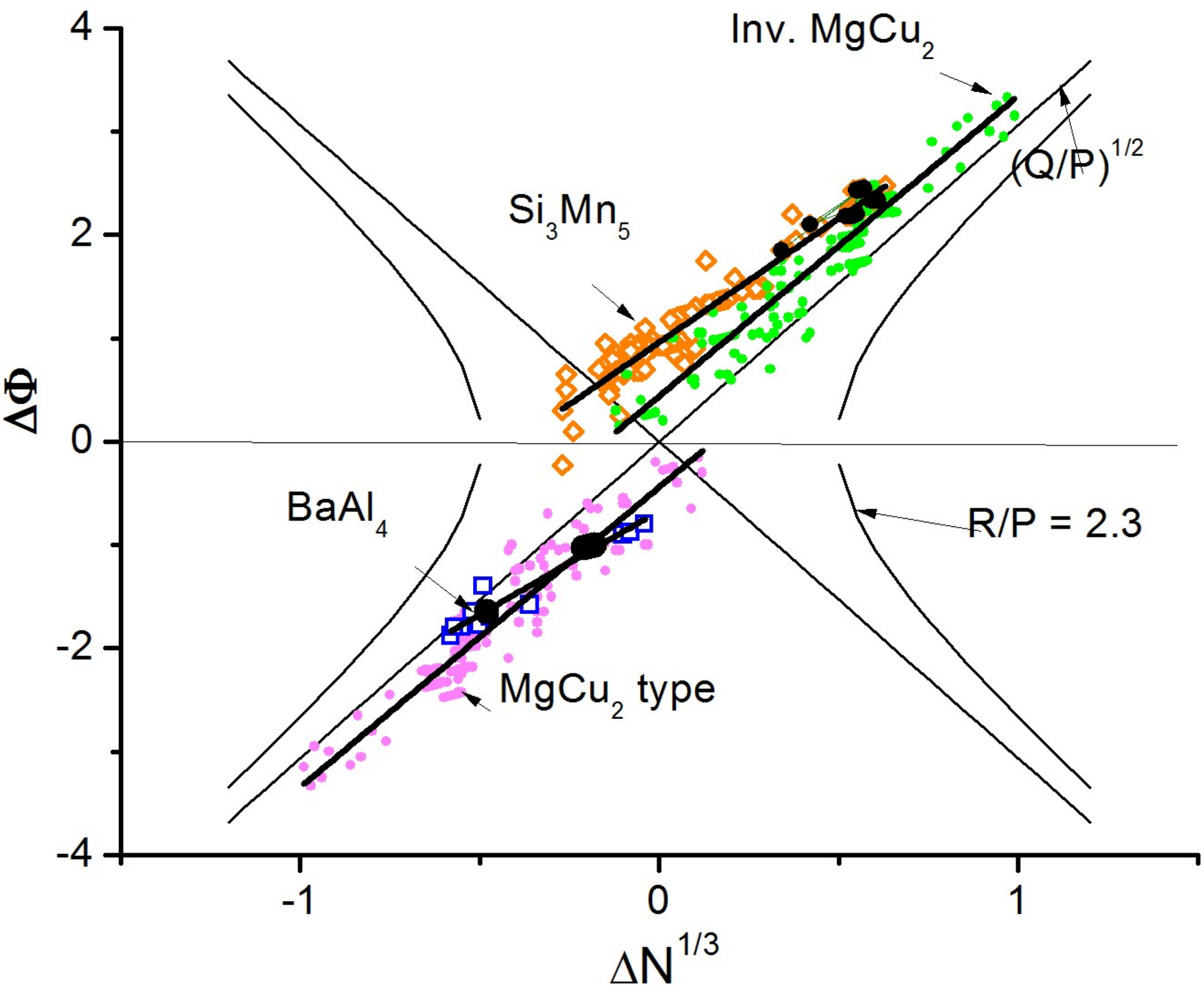

**FIG. 2.** (colour online) The behaviour on a $(\Delta\phi,\ \Delta N^{1/3})$ map of the $RG$ lines of some structure types which are concomitant with the $MgCu_2$ type is demonstrated. The figure demonstrates the ability of $RG$ lines to predict concomitant structure types. The $RG$ line of $BaAl_4$ type and that of $MgCu_2$ type overlap in a region and the 7 binary systems that have both the structure types (at $AB_4$ and $AB_2$ compositions respectively) are seen as circular agglomeration of points, black online. We note that there is no binary system which has both the structure types outside the region of overlap anywhere in the map. $BaAl_4$ compounds at composition $AB_4$ are never accompanied by $MgCu_2$ compounds at composition $A_2B$ in the same binary system. We also see from the figure that the 24 binary systems in which both $MgCu_2$ and $Si_3Mn_5$ type structures are concomitant occur only in the region where the RG lines of the two structure types intersect (seen as large circular points, black online).

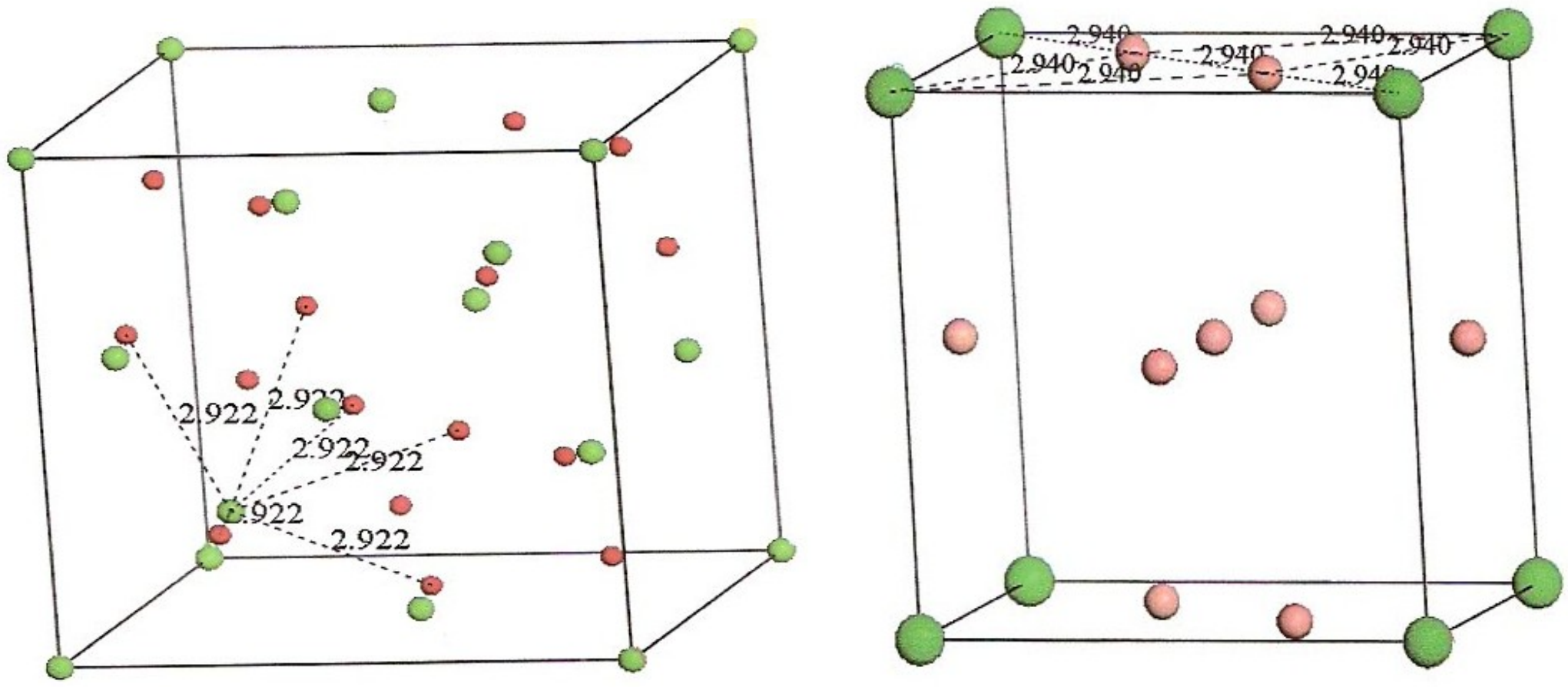

**Fig. 3** (colour online) **(a)** The unit cell of $MgCu_2$ type ($AB_2$, cF24) compounds. *A* atoms are shown larger in diameter (green online). There are 12 bonds of equal length to the *B* atoms from each *A* atom. In the compound $MgCu_2$, the Mg-Cu nearest neighbour distance is 2.922 Å [30]. **(b)** The unit cell of $CaCu_5$ type ($AB_5$, hP6) compounds. *A* atoms are shown bigger (green online). There are 6 bonds of equal length to the *B* atoms from each *A* atom. In the compound $CaCu_5$, the nearest neighbour Ca-Cu distance is 2.940 Å[32].